\documentclass[aps,preprint,prd,showpacs,nofootinbib]{revtex4-2}

\usepackage{amsmath,amssymb}
\usepackage{graphicx,subfigure}
\usepackage{color,multirow}
\usepackage[colorlinks,linkcolor=magenta,anchorcolor=cyan,citecolor=blue,plainpages=false]{hyperref}

\usepackage{caption}

\usepackage{CJKutf8}

\newcommand{\apjl}{Astrophys. J. Lett.}
\newcommand{\mnras}{Mon. Not. R. Astron. Soc.}
\newcommand{\apjs}{Astrophys. J. Suppl. Ser.}
\newcommand{\sovast}{Sov. Astron.}
\newcommand{\aap}{Astron. Astrophys.}
\newcommand{\jcap}{JCAP}
\newcommand{\aj}{Astron. J.}
\newcommand{\pasp}{Publ. Astron. Soc. Pac.}

\begin{document}

\title{Influence of supermassive primordial black holes on ultraviolet luminosity of high-redshift galaxies}

\author{Yi-Yang Li$^{1,2} $\footnote{\href{liyiyang20@mails.ucas.ac.cn}{liyiyang20@mails.ucas.ac.cn}}}
\author{Hai-Long Huang$^{1,2} $\footnote{\href{huanghailong18@mails.ucas.ac.cn}{huanghailong18@mails.ucas.ac.cn}}}
\author{Yun-Song Piao$^{1,2,3,4} $ \footnote{\href{yspiao@ucas.ac.cn}{yspiao@ucas.ac.cn}}}

    \affiliation{$^1$ School of Fundamental Physics and Mathematical
        Sciences, Hangzhou Institute for Advanced Study, UCAS, Hangzhou
        310024, China}

    \affiliation{$^2$ School of Physics Sciences, University of
        Chinese Academy of Sciences, Beijing 100049, China}

    \affiliation{$^3$ International Center for Theoretical Physics
        Asia-Pacific, Beijing/Hangzhou, China}

    \affiliation{$^4$ Institute of Theoretical Physics, Chinese
        Academy of Sciences, P.O. Box 2735, Beijing 100190, China}


\begin{abstract}
Recently James Webb Space Telescope (JWST) have observed an excess
of luminous galaxies at high redshifts ($z \gtrsim 10$). In this
work, we investigate whether supermassive primordial black holes
(SMPBHs) can explain it by their influence on the ultraviolet
luminosity function (UV LF) of high-redshift galaxies. 
Through Markov Chain Monte Carlo analysis, we constrain the parameters
relevant with SMPBHs against current JWST observational data. The
results reveal that SMPBHs with masses $M_{\rm PBH} \sim
10^{6.3\text{-}8.3} M_\odot$, abundances $f_{\rm PBH} \sim
10^{-7}\text{-}10^{-5}$, and sub-Eddington ratios $\lambda_E \ll
1$ can effectively enhance the bright end of the UV LF, consistent
with JWST observations.


\end{abstract}


\maketitle

\section{Introduction} \label{sec:intro}

The James Webb Space Telescope (JWST) has opened an unprecedented
observational window into the early Universe, enabling detailed
investigations of the birth of the first galaxies and stars. JWST
have uncovered a surprisingly abundant population of massive
galaxies at high redshifts ($z \gtrsim 10$), significantly
exceeding the predictions of standard galaxy formation model
\citep{Labbe_2023,Robertson_2023,Robertson_2024,Casey_2023,P_rez_Gonz_lez_2023}.
In particular, the observed excess at the bright end of the
ultraviolet luminosity function (UV LF) presents a notable
challenge.
\citep{Mcleod_2023,Bouwens_2023a,Adams_2024,Harikane_2023,
Harikane_2024,Finkelstein_2023a,Finkelstein_2023b,Donnan_2023,
Donnan_2024}
Reconciling these observations with standard models would require
an unrealistically high star formation efficiency (SFE)
\citep{Bouwens_2023b,Oesch_2018},
or alternatively, explaining the data within the $\Lambda$CDM
framework by allowing the SFE to evolve with redshift
\citep{2023ApJ...954L..48W}.

It is intriguing to consider the potential role of primordial
black holes (PBHs)
\citep{1974MNRAS.168..399C,1967SvA....10..602Z,1971MNRAS.152...75H,
1975ApJ...201....1C,1975Natur.253..251C,1993PhRvD..48..543C,
1994PhRvD..50.7173I,1996PhRvD..54.6040G,1996NuPhB.472..377R,
2019PhRvL.123g1102M,2016PhRvL.116t1301B,2016PhRvL.117f1101S,
2001JETP...92..921R,2002astro.ph..2505K,2005APh....23..265K,
Khlopov_2007,2010RAA....10..495K,2019EPJC...79..246B,2024SCPMA..6711011G},
which have been extensively studied as viable candidates for dark
matter, especially supermassive PBHs (SMPBHs) with $M_{\rm
PBH}\gtrsim 10^6M_\odot$. PBHs, especially SMPBHs, 
might be explanations for the
supermassive galaxies\citep{2024SCPMA..6709512Y,Huang_2024c,Huang_2024b} and "Little
Red Dots" (LRDs)\citep{Huang_2024a} recently discovered by the
JWST. It has been showed that accreting PBHs can contribute
additional ultraviolet luminosity, boosting the number density of
bright galaxies \citep{Matteri_2025a,Matteri_2025b}, while the
Poisson fluctuations in spatial distribution of PBHs can modulate
the halo mass function (HMF), enhancing the abundance of collapsed
structures at high redshift \citep{Liu_2022}, see also the impact
of initial clustering of SMPBHs \citep{Huang_2024b}.

These SMPBHs could help to bridge the gap between the standard
$\Lambda$CDM model and JWST observations. The mass range of PBHs
can extend from $10^{-18}\rm~M_\odot$ to $10^{16}\rm~M_\odot$
\footnote{Those smaller than $\sim10^{-18}\rm~M_\odot$ would have
evaporated by now due to Hawking radiation, or see recent
\citep{2025PhRvD.111b4009C,2025PhRvD.111b4010C,
Calza:2025mwn}.},
although current observations indicate that SMBHs in galactic
nuclei typically have masses extending up to only nearly
$10^{11}\rm~M_\odot$ \citep{2011Natur.480..215M}. In the standard
scenario that PBHs were sourced by the collapse of large density
fluctuations
\citep{2016PhRvD..94h3504C,Garcia-Bellido:2017fdg,2020ARNPS..70..355C,
2021RPPh...84k6902C,2021JPhG...48d3001G,Carr:2021bzv,
2024bheg.book..261E,2018CQGra..35f3001S,2024CQGra..41n3001D},
massive PBHs with $M\gtrsim 10^5M_\odot$ is strictly constrained
by CMB $\mu$-distortion observations
\citep{2014PhRvD..90h3514K,2023PhRvL.130q1401D,
2018PhRvD..97d3525N,2021PhRvD.104f3526D}. However, it can be
expected that significant non-Gaussianity of primordial
perturbations \citep{2021PhRvD.103f3519A,
2023EL....14249001G,2024JCAP...04..021H,2024JCAP...09..012B,
2024JCAP...07..090S,2024PhRvL.133b1001C,2024JCAP...10..050I,
Wang_2025,2025JCAP...07..079P} 
might be helpful for overcoming this difficulty, or alternative
mechanisms, such as PBHs that the supercritical bubbles nucleated
during inflation developed into (which are not constrained by
$\mu$ distortion, e.g.\citep{2024JCAP...09..012B}), must be
considered. In the latter case, the resulting SMPBHs can have a
(multiple) peak-like mass spectrum
\citep{Huang:2023chx,2024PhRvD.110b3501H}.
Therefor, it is significant to further investigate the role of
SMPBHs in shaping the early Universe.

This work focuses on how SMPBHs can influence the UV LF of
high-redshift galaxies through both Poisson effects and
accretion-induced emission, potentially offering an explanation
for the recent JWST observations. The values of cosmological
parameters adopted in this paper are the bestfit values of Planck
based on the $\Lambda$CDM model
\citep{Placnck_2020}\footnote{Here, we do not consider the
potential impact of the resolution of Hubble tension on the
bestfit values of relevant cosmological parameters,
e.g.\cite{2020PhRvD.101h3507Y,2021PhRvD.104f3510Y,
2022PhRvD.105j3514J,2024MNRAS.527L..54J,
2024PhLB..85138588J,Wang:2024dka,Wang:2024tjd},
which might be actually negligible. }.

\section{Influence of SMPBHs on UV luminosity}

It has been showed in Ref.\citep{Huang:2023chx} that in
string landscape scenario with multiple metastable vacua (usually
modelled as a multi-dimensional potential), when the inflaton
slowly passes by a neighboring vacuum, the nucleating rate of
supercritical bubbles would inevitably present a peak, so that the
resulting PBHs have a non-negligible distinctive peak-like mass
distribution \citep{Huang_2023}:
\begin{equation}\label{eq:psi_m1}
\psi_{1}(m) = e^{-\sigma^2/8} \cdot \sqrt{\frac{M_{PBH}}{2\pi
\sigma^2 m^3}} \cdot \exp\left[ -\frac{\ln ^2 (m /
M_{PBH})}{2\sigma^2} \right],
\end{equation}
where $M_{PBH}$ is the characteristic mass of PBHs, and $\sigma$
is the width of mass peak. 
This power-law term causes the distribution to go down
more rapidly at larger masses. The corresponding mean mass of the
distribution is: \( \langle m \rangle = M_{PBH} e^{-\sigma^2} \)

Here, a conversion relation is required to connect
Eq.~(\ref{eq:psi_m1}) with the number-density-related mass
function of PBHs:
\begin{equation}\label{eq:psi_m3}
\psi_{3}(m) = \frac{\langle m\rangle}{m} \psi_{1} =
e^{-9\sigma^{2}/8} \sqrt{\frac{M_{PBH}^{3}}{2\pi\sigma^{2}m^{5}}}
\exp\left[-\frac{\ln^{2}(m/M_{PBH})}{2\sigma^{2}}\right].
\end{equation}
This distribution differs from the
standard log-normal distribution by including an extra $m^{-3/2}$
factor.
The interrelations between various definitions of mass
distribution and the corresponding expectation values are
summarized in Table~\ref{tab:S1rules}. Thus we have
\begin{equation}\label{eq:dn_dm_PBH}
\frac{dn}{n_0 dm} = \psi_3,
\end{equation}
with the comoving number density of PBHs $n_{0} = \frac{
\rho_{c}\Omega_{dm}f_{\mathrm{PBH}}}{\langle m\rangle}   =
\frac{\rho_{c}\Omega_{dm}f_{\mathrm{PBH}}} {M_{PBH}}
e^{\sigma^{2}}$, where $n$ is the number density of PBHs,
\(\Omega_{dm} \) is the dark matter density parameter and \(
\rho_{c} \) is the critical density  of the Universe.


The SMPBHs sourced by such inflationary vacuum bubbles can avoid
constraint from spectral distortions in the CMB, and thus can be
supermassive, $M_{PBH}\gtrsim 10^6M_\odot$. This mechanism offers
a natural route to SMPBHs coming into being without requiring
large curvature perturbations.

It can be expected that SMPBHs would enhance galaxy luminosity
through the accretion of matter and subsequent emission of
radiation. This process could potentially boost the bright end of
the UV LF. Here, we quantify the contribution of SMPBHs to the
luminosity using the Eddington ratio, $\lambda_E$. Specifically,
the bolometric luminosity of a PBH is\footnote{Here, all PBHs are
assumed to have the same Eddington ratio \(\lambda_E\), although
in reality \(\lambda_E\) may follow a broad distribution. The case
of \(\lambda_E = 0\) corresponds to considering only the Poisson
     effect.} $L_{\mathrm{bol}}(m_{\mathrm{PBH}}) = \lambda_E
L_E(m_{\mathrm{PBH}})$. Here, $L_E$ represents the Eddington
luminosity, defined as:
\begin{equation}
L_E(m_{\mathrm{PBH}}) = \frac{4\pi G m_p c}{\sigma_T}
m_{\mathrm{PBH}} = 3.28 \times 10^4
\frac{m_{\mathrm{PBH}}}{M_{\odot}} L_{\odot},
\end{equation}
where $m_p$ denotes the proton mass, $c$ is the speed of light,
$\sigma_T$ signifies the Thomson scattering cross-section,
$m_{\mathrm{PBH}}$ is the mass of PBHs, and solar Luminosity is
$L_{\odot} \approx 3.828 \times 10^{33} \, \text{erg s}^{-1}$.

According to the prescription of \citet{Shen_2020}, the bolometric
luminosity can be converted into the UV luminosity,
\begin{equation}\label{eq:LUV-PBH}
L_{\mathrm{UV, PBH}}(m_{\mathrm{PBH}}) =
\frac{L_{\mathrm{bol}}(m_{\mathrm{PBH}})}{c_1 \left(
\frac{L_{\mathrm{bol}}(m_{\mathrm{PBH}})}{10^{10} L_{\odot}}
\right)^{k_1} + c_2 \left(
\frac{L_{\mathrm{bol}}(m_{\mathrm{PBH}})}{10^{10} L_{\odot}}
\right)^{k_2}},
\end{equation}
where
\[
\begin{pmatrix} c_1 \\ k_1 \\ c_2 \\ k_2 \end{pmatrix} = \begin{pmatrix} 1.862 \\ -0.361 \\ 4.870 \\ -0.0063 \end{pmatrix}.
\]
The UV luminosity \(L_{\mathrm{UV,\,PBH}}\) here refers to the
band UV luminosity of AB magnitude system measured in a top-hat
filter centered at rest-frame 1450\,\AA\ with a bandwidth of
100\,\AA, or equivalently, the luminosity \(\nu L_\nu\) at 1450.



\section{Confronting SMPBH model with JWST observations}\label{sec:methods}

\subsection{Total UV luminosity with stars and
PBHs}


The UV LF:
\begin{equation}\label{eq:UV_LF}
\Phi_{\rm UV} \equiv \frac{dn}{dM_{\rm UV}}(z,M_{\rm UV})
\end{equation}
quantifies the number density of galaxies per luminosity or
absolute UV magnitude interval. It can be related to the dark
matter HMF by using $\Phi_{\rm UV}= {\left|\frac{dM_{\rm
UV}}{dM_{\rm halo}}\right|}^{-1}\frac{dn}{dM_{\rm halo}}$, where
the HMF $\frac{dn}{dM_{\rm halo}}$ can be calculated with the Sheth-Tormen
algorithm \citep{Sheth_1999}.

The monochromatic UV luminosity and AB magnitude are related by
\citep{Oke_1983}
\begin{equation}\label{eq:MUV-LUV}
M_{\rm UV} = -2.5\log\left(\frac{L_{UV}/
\nu}{\mathrm{erg\,s^{-1}\,Hz^{-1}}}\right) + 51.60.
\end{equation}
In this work, we consider the ultraviolet wavelength band at \(
\nu = 1450\,\text{\AA} \). It is important to note whether the
luminosity used is the monochromatic UV luminosity \( L_\nu \) or
the band UV luminosity \( L_{\mathrm{UV}} \), both are
approximately related by $L_{\mathrm{UV}} \approx \nu  L_\nu$.

\begin{figure}[htbp]
    \begin{minipage}[b]{0.8\textwidth}
        \centering
        \includegraphics[width=\textwidth]{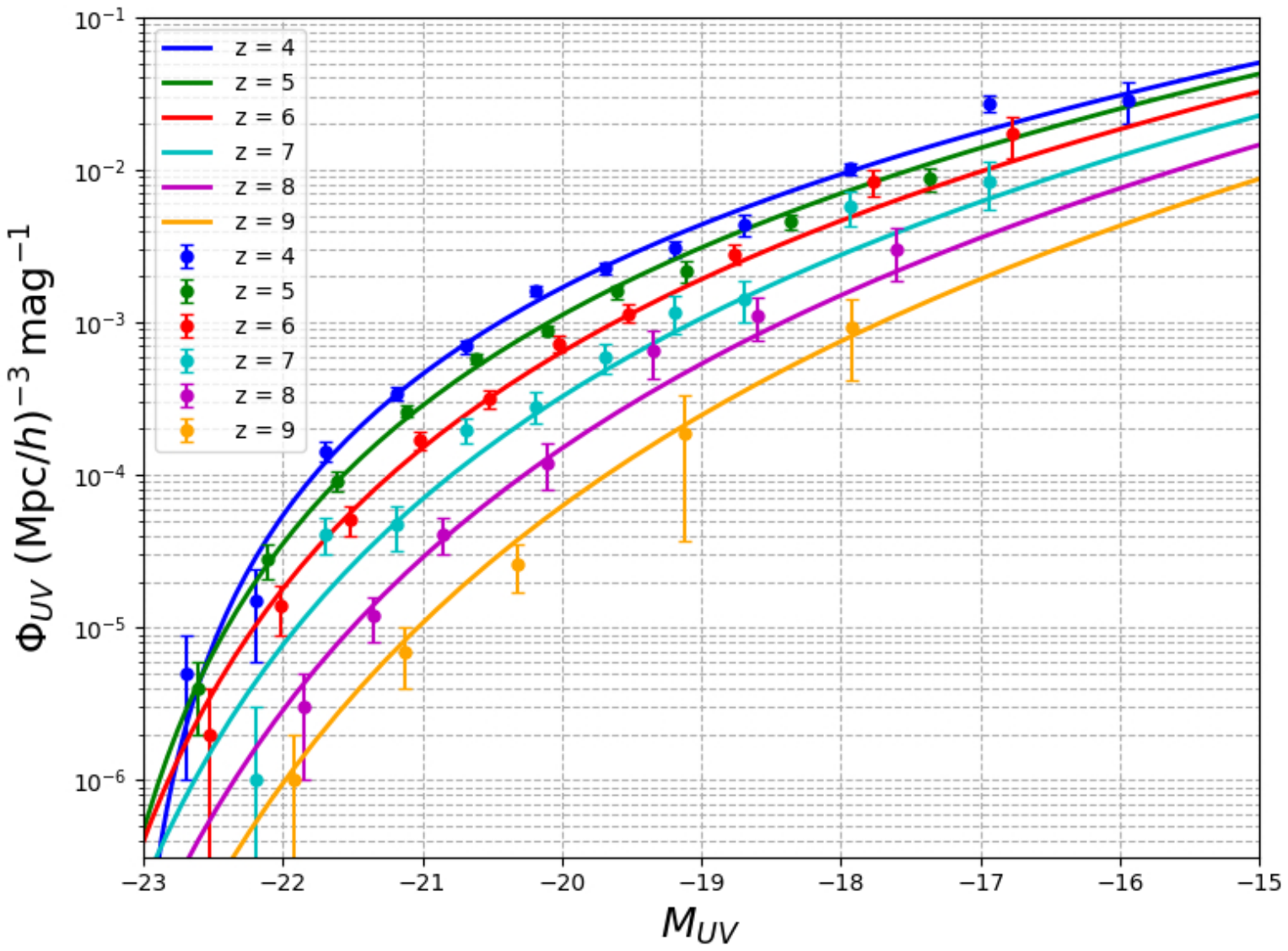}
    \end{minipage}
\caption{
    The corresponding UV LF
    in Eq.~(\ref{eq:mh_muv_fit}) with best-fit parameters (\ref{eq:fit-results-muv-vs-mh}),
    in comparison with observational data at
    redshifts $z = 4$-$9$\citep{Bouwens_2021}.
    }
    \label{fig:1}
\end{figure}

Here, we adopt a purely empirical fit of the $M_{\rm
halo}$-$M_{\rm UV}$ relation \citep{Matteri_2025b}:
\begin{equation}\label{eq:mh_muv_fit}
M_{\rm UV}= p_1 + \left[ p_2 \left(\frac{M_{\rm halo}}{10^{11}
M_\odot}\right)^{p_3} + p_4\right] \log \left(\frac{M_{\rm
halo}}{10^{11} M_\odot}\right) + p_5 z^{p_6}
\end{equation}
where $p_i$ are set
with a least-squares fit of Eq.~(\ref{eq:mh_muv_fit}) to the
observed UV LF data \citep{Bouwens_2021} (see Figure~\ref{fig:1}),
\begin{equation}\label{eq:fit-results-muv-vs-mh}
    \begin{pmatrix}
        p_1 \\
        p_2 \\
        p_3 \\
        p_4 \\
        p_5 \\
        p_6
    \end{pmatrix}_{\rm fit}
    =
    \begin{pmatrix}
        -13.12 \\
        -8.94 \\
        -0.0355 \\
        5.78 \\
        -3.84 \\
        0.341
    \end{pmatrix}
\end{equation}



The total UV luminosity can be contributed by not only stars but
also SMPBHs, when we consider the impact of SMPBHs, i.e.,
\begin{equation}\label{eq:LUV-tot}
L_{\mathrm{UV,\,tot}}(M_{\rm halo},z,m_{\mathrm{PBH}}) =
L_{\mathrm{UV}}(M_{\rm halo}, z) +
L_{\mathrm{UV,\,PBH}}(m_{\mathrm{PBH}}),
\end{equation}
where the stellar UV magnitude \(M_{\mathrm{UV}}\)
(\ref{eq:mh_muv_fit}) can be converted to the UV luminosity
\(L_{\mathrm{UV}}(M_{\rm halo}, z)\) using Eq.~(\ref{eq:MUV-LUV}).


\subsection{Method for performing MCMC analysis}\label{sec:Numerical Methods}


The number distribution of PBHs within dark matter halo follows a
Poisson distribution with mean $\lambda$,
\begin{equation}
\label{Poisson} \lambda = \frac{n_0}{n_{\mathrm{halo}}} =
\frac{f_{\mathrm{PBH}} \Omega_{dm} \rho_c}{n_{\mathrm{halo}}
\langle m_{\mathrm{PBH}} \rangle} = \frac{f_{\mathrm{PBH}}
\Omega_{dm} \rho_c}{n_{\mathrm{halo}} M_{\mathrm{PBH}}}
e^{\sigma^2},
\end{equation}
where\footnote{Here, we have assumed that PBHs are only
distributed in dark matter halos with $M_{\rm halo}>M_{\min}$. Due
to the exponential span in mass, low-mass halos only significantly
affect the case that the luminosity from PBHs far exceeds the
stellar component. } \[
n_{\mathrm{halo}}=\int_{M_{min}(z)}^{\infty} \frac{dn}{dM_h}(M_h,
z) dM_h \,
\] is the comoving number density of
halos.

In light of the mass function \eqref{eq:dn_dm_PBH} of PBHs and the
luminosity model \eqref{eq:LUV-PBH}, we can have the luminosity
distribution of a single PBH whose number density follows a
Poisson distribution, see Eq.(\ref{Poisson}).
The sum of luminosities of all PBHs follows a compound Poisson
distribution. Therefore, the total luminosity function contributed
by stars and PBHs is the convolution of the stellar luminosity
distribution and the compound Poisson distribution of PBHs.

Here, we will perform the Markov Chain Monte Carlo (MCMC) analysis
with observational data. The \texttt{emcee} package
\citep{emcee_2013} was employed to sample the parameters $\log
M_{\mathrm{PBH}}$, $\log f_{\mathrm{PBH}}$, and $\sigma$ while
fixing the Eddington ratio $\lambda_E=0.01$, $0.1$ and $1$,
respectively.

\begin{table}[htbp]
\centering
\begin{tabular}{|c|c|}
\hline
\textbf{Parameter} & \textbf{Range} \\
\hline
$\log f_{\mathrm{PBH}}$ & $[-10, -3]$ \\
\hline
$\log \left( M_{\mathrm{PBH}} / M_\odot \right)$ & $[1, 11]$ \\
\hline
$\sigma$ & $[0.1, 2.5]$ \\
\hline
\end{tabular}
\caption{The flat prior of relevant parameters $\log
f_{\mathrm{PBH}}$,
    $\log M_{\mathrm{PBH}}/M_\odot$ and $\sigma$.}
\end{table}

Accounting for the range of the luminosity function, the
log-likelihood function is defined as:
\[
\ln \mathcal{L} = -\frac{1}{2} \sum_{i} \left( \frac{\log \phi_{\mathrm{model},i} - \log \phi_{\mathrm{obs},i}}{\sigma_{\log,i}} \right)^2
\]
where $\phi_{\mathrm{model},i}$ is the model prediction at
magnitude $M_{UV,i}$, $\phi_{\mathrm{obs},i}$ is the observed
value, and the logarithmic error is $\sigma_{\log,i} = \sigma_i /
(\phi_i \ln 10)$.

\subsection{Results}\label{sec:result}

\begin{figure}[!htbp]
    \centering
    \includegraphics[width=0.8\textwidth]{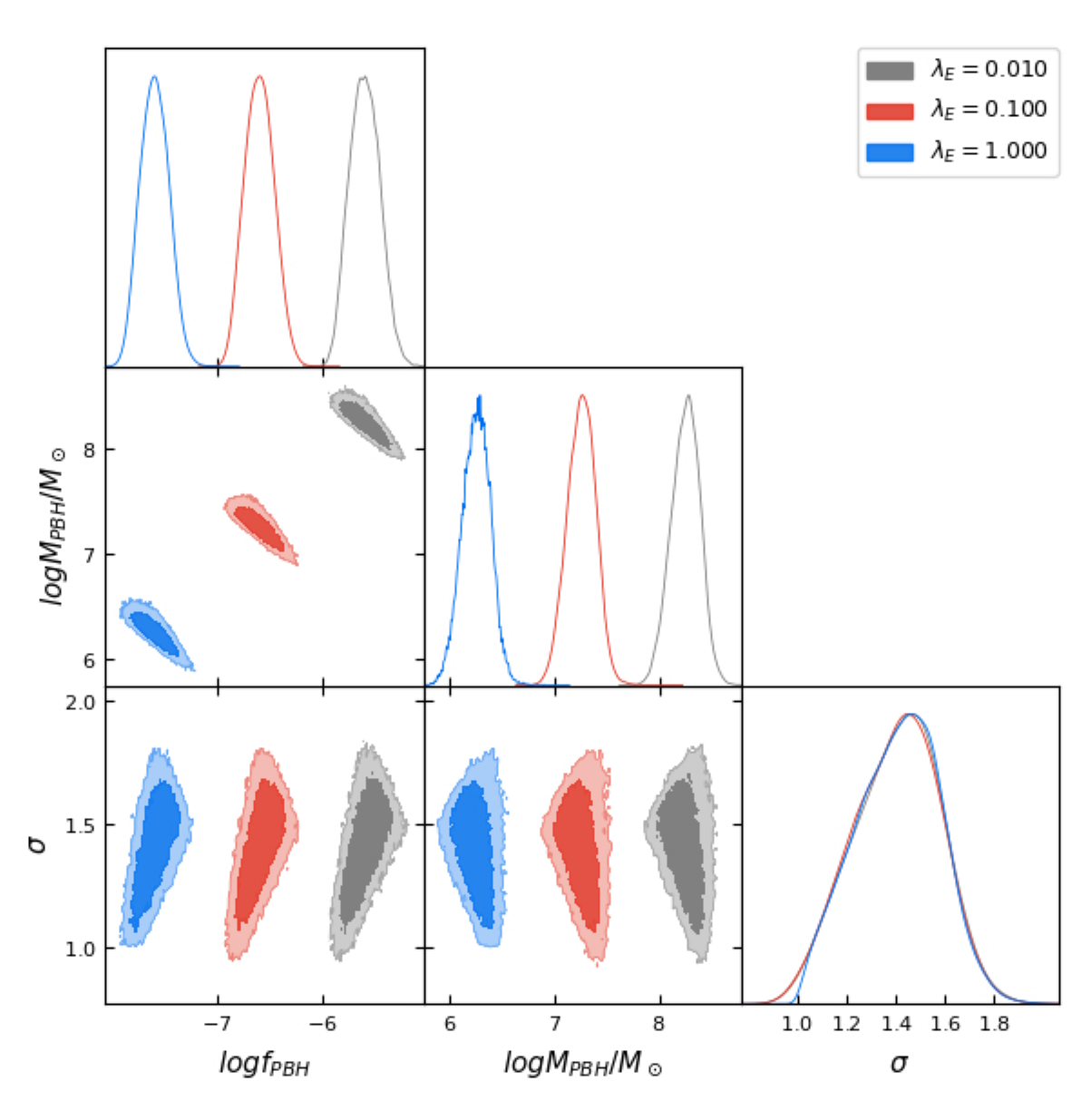}
    \caption{The MCMC posterior distributions at
    redshift $z=11$ for fixed Eddington ratios $\lambda_E = 0.100$ (red) ,
    $\lambda_E = 1.000$ (blue) and $\lambda_E = 0.010$ (gray).}
    \label{fig:3}
\end{figure}

The relevant model parameters are \(f_{\rm PBH}\), \(M_{\rm
PBH}\), \(\lambda_{E}\), and \(\sigma\). In
Table~\ref{tab:lambdaE} and Figure~\ref{fig:3}, we present the
posterior distributions of corresponding parameters.

The mass function \eqref{eq:psi_m3} of PBHs has an additional mass
correction factor of \( m^{-1.5} \) compared to the standard
log-normal distribution. This results in a faster decay of the
probability distribution at the high-mass end, necessitating
relatively larger PBH mass parameters \( M_{\mathrm{PBH}} \) (with
\( \lambda_E = 0.1 \), \( M_{\mathrm{PBH}} = 10^{7.26} M_\odot
\)).

\begin{table}[htbp]
    \centering
    \begin{tabular}{cccc}
        \hline
        $\lambda_E$ & $\log f_{\mathrm{PBH}}$ & $\log M_{\mathrm{PBH}}/M_\odot$ & $\sigma$ \\
        \hline
        0.010 & $-5.60^{+0.15}_{-0.14}$ & $8.26^{+0.13}_{-0.15}$ & $1.42^{+0.17}_{-0.20}$ \\
        0.100 & $-6.59^{+0.14}_{-0.14}$ & $7.26^{+0.13}_{-0.14}$ & $1.41^{+0.17}_{-0.20}$ \\
        1.000 & $-7.59^{+0.14}_{-0.14}$ & $6.26^{+0.13}_{-0.14}$ & $1.42^{+0.16}_{-0.19}$ \\
        \hline
    \end{tabular}
    \caption{Posterior constraints on the model parameters for
    different fixed Eddington ratios $\lambda_E$,
    derived from MCMC sampling.
    The table lists the median values and 68\% credible intervals
    for $\log f_{\mathrm{PBH}}$,
    $\log M_{\mathrm{PBH}}/M_\odot$, and $\sigma$.
    }
    \label{tab:lambdaE}
\end{table}

In Figure~\ref{fig:3}, a positive correlation between peak values
of $f_{\mathrm{PBH}}$ and $M_{\mathrm{PBH}}$ is revealed, which
physically originates from the requirement that the number density
of PBHs is approximately constant. This conserved number density
ensures statistically significant enhancement of the UV LF. In the
vicinity of peak, $M_{\mathrm{PBH}}$ and $f_{\mathrm{PBH}}$
exhibit a negative correlation, while a corresponding negative
correlation exists between peak $M_{\mathrm{PBH}}$ and
$\lambda_E$:
\[
\label{M-lambda} M_{\mathrm{PBH}} \lambda_E \approx 10^{6.26}
M_\odot.
\]
This arises because, under relatively constant number density
conditions, $L_{\mathrm{PBH}}$ maintains a positive correlation
with both $M_{\mathrm{PBH}}$ and $f_{\mathrm{PBH}}$. It can be
also seen that lower PBH masses necessitate broader mass
distribution widths to offer sufficient massive PBHs that impact
the observational data range, notably the width parameter $\sigma$
shows nearly no changes across different Eddington ratios.

\begin{figure}[htbp]
    \centering
    \includegraphics[width=0.8\textwidth]{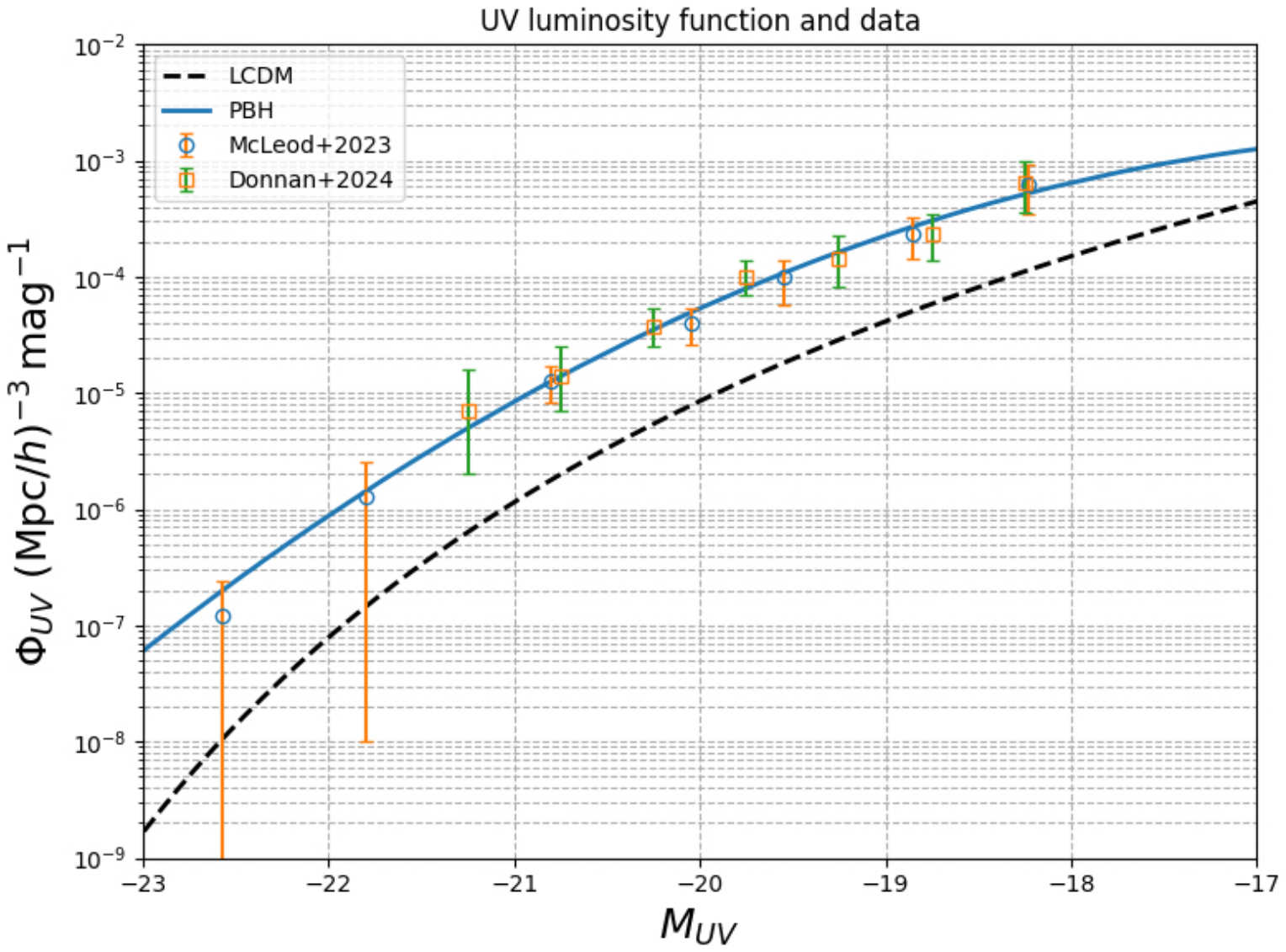}
\caption{Influence of SMPBHs with the mass spectrum
(\ref{eq:psi_m3}) on the UV LF at \(z=11\), where \(\lambda_E =
0.1\).
    The model parameters are labelled in the Table~\ref{tab:lambdaE}.
    Data points from \citet{Mcleod_2023}(blue circles)
    and \citet{Donnan_2024}(orange squares) are also shown for comparison.
    Compared to the standard \(\Lambda\)CDM result (black dashed line),
    the UV LF from the \(\Lambda\)CDM+SMPBH model is notably boosted by both the
    Poisson effect and the intrinsic UV emission of SMPBHs (blue solid line).
}
    \label{fig:4}
\end{figure}

Therefore our result (\ref{M-lambda}) suggests that the PBHs must
be supermassive for $\lambda_E<1$. In Figure~\ref{fig:4}, we
compare the UV LF of the \(\Lambda\)CDM+SMPBH model with the
standard \(\Lambda\)CDM scenario at \(z=11\). The results show
that the inclusion of SMPBHs, especially with accretion emission,
leads to a significantly improved fit to the observed data at the
bright end \citep{Mcleod_2023,Donnan_2024}. This demonstrates that
our SMPBHs model can account for the excess of luminous galaxies
detected by JWST.



\section{DISCUSSION}\label{sec:discussion}

In this work, we explore how SMPBHs in different mass ranges,
sourced by inflationary vacuum bubbles, can influence the UV LF of
high-redshift galaxies. It is found that the high-redshift galaxy
UV LF can be notably boosted by the intrinsic UV emission of SMPBHs
($M\gtrsim 10^6M_\odot$), with the bright-end excess primarily
driven by sub-Edington accreting SMPBHs, see (\ref{M-lambda}).
The bright-end enhancement observed in Figure~\ref{fig:4} can
originate from the high-mass tail of the contribution
(\ref{eq:psi_m3}) of SMPBHs, which requires the distribution to
have a relatively large width parameter $\sigma$.



However, our discussion has not yet considered the possible mass
growth of PBHs with redshift. It seems that if this effect is
taken into account, the corresponding initial mass for PBHs would
be relatively smaller, however, for SMPBHs the case might be
different, the SMPBHs $(\gtrsim 10^6M_\odot)$ can follow the
self-similar accretion, so that their mass growth during
pregalactic era might be negligible, e.g.\citep{Huang_2024c}.
According to our results, SMPBHs sourced by the inflationary
vacuum bubbles are very likely to be regarded as SMBHs when the
redshift $z\lesssim 20$, which can accelerate seeding
high-redshift galaxies and their baryonic content. It will be
interesting to further explore precise constraints on the mass
distribution parameters (such as the characteristic mass $M_{PBH}$
and width $\sigma$) of SMPBHs with more upcoming observations from
JWST.


\begin{acknowledgments}

This work is supported by National Key Research and Development
Program of China, No.2021YFC2203004, and NSFC, No.12475064, and
the Fundamental Research Funds for the Central Universities.

\end{acknowledgments}

\vspace{5mm}

\newpage





\appendix

\section{Definitions of normalized PBH mass distributions in different references}

\begin{table}[!ht]
    \centering
    \begin{tabular}{|c|c|c|c|}
       \hline
        \citet{Escriv_2023} & \citet{Raidal_2024} &  &  \\
        \citet{Clesse_2021}  & \citet{Franciolini_2022} & \citet{Hutsi_2021} & \citet{Kocsis_2018} \\
        \citet{Bagui_2023b} & \citet{Hall_2020} &  & \\
        \hline
        $f\equiv \dfrac{1}{\rho_{\rm PBH}} \dfrac{{\rm d} \rho_{\rm PBH} }{{\rm d}\ln m}   $  & $ \psi_{1} \equiv \dfrac{1}{\rho_{\rm PBH}} \dfrac{{\rm d} \rho_{\rm PBH} }{{\rm d} m}   $  & $ \psi_{2} \equiv \dfrac{1}{n_{\rm PBH}} \dfrac{{\rm d} n_{\rm PBH} }{{\rm d} \ln m}   $ & $ \psi_{3}\equiv \dfrac{1}{n_{\rm PBH}} \dfrac{{\rm d} n_{\rm PBH} }{{\rm d}  m}   $ \\
        \hline
        $f =  m \psi_1  $ & $\psi_1 = f/m $ & $ \psi_2 = f \langle m \rangle /m$   & $ \psi_3= \langle m \rangle f / m^2  $  \\
        $ = m \psi_2 / \langle m \rangle $ & $= \psi_2 / \langle m \rangle $  & $ = \langle m \rangle \psi_1$ & $  = \langle m \rangle \psi_1 /m $ \\
        $ = m^2 \psi_3 / \langle m \rangle   $  & $ =  m \psi_3 / \langle m \rangle   $  & $= m \psi_3 $  & $ = \psi_2/m $ \\
       \hline
       $\int f {\rm d} \ln m = 1$ & $\int \psi_1 {\rm d} m = 1$ & $\int \psi_2  {\rm d} \ln m = 1$ & $ \int \psi_3 {\rm d} m = 1$  \\
        \hline
       $\langle m \rangle = \left( \int \dfrac{f}{m} {\rm d} \ln m \right)^{-1} $  & $\left( \int \dfrac{\psi_1}{m} {\rm d} m \right)^{-1}$  & $ \int m \psi_2 {\rm d} \ln m $ & $ \int m \psi_3 {\rm d} m $ \\
       \hline
       $ \langle m^2 \rangle = \langle m \rangle \int  m f {\rm d} \ln m $ & $ \langle m \rangle \int m \psi_1 {\rm d } m $   &  $ \int m^2 \psi_2 {\rm d} \ln m  $ &  $ \int m^2 \psi_3 {\rm d}{m } $ \\
       \hline
    \end{tabular}
\caption{Definitions of normalized PBH mass distributions proposed
in different literatures, along with their interconversion
relations, normalization rules, and corresponding expectation
values $\langle m \rangle$ and $\langle m^2 \rangle$.
The table is adapted from \citet{Escriv_2023}, presented in
\citet{Bagui_2023a}.
    }
    \label{tab:S1rules}
\end{table}












\section{Influence of the PBH mass distribution on the power spectrum}\label{app:pbh_mass_function_power}

The influence of the Poisson effect \citep{2018MNRAS.478.3756C} induced by
PBHs with a monochromatic mass distribution on the matter power
spectrum is \citep{Inman_2019}:
\begin{equation}\label{eq:P_tot}
    P_{\mathrm{tot}} = P_{\Lambda \mathrm{CDM}} + P_{\mathrm{PBH}},
\end{equation}
with
\begin{equation}\label{eq:P_PBH}
    P_{\mathrm{PBH}} = \frac{f_{\mathrm{PBH}}^2}{n_{\mathrm{PBH}}} D_0^2 \, \Theta(k_{\mathrm{cut}} - k) \qquad for \quad k>k_{eq}
\end{equation}
The PBH contribution depends on the PBH dark matter fraction \(
f_{\mathrm{PBH}} \) and the comoving number density
\[
n_{\mathrm{PBH}} = \frac{f_{\mathrm{PBH}}}{M_{\mathrm{PBH}}}
\left( \Omega_{\mathrm{m}} - \Omega_{\mathrm{b}} \right) \frac{3
H_0^2}{8 \pi G}
\]
where $M_{\mathrm{PBH}}$ is the mass of PBHs,
$\Omega_{\mathrm{m}}$ represents the total matter density
parameter (including both dark and baryonic matter),
$\Omega_{\mathrm{b}}$ denotes the baryonic matter density
parameter, $H_0$ is the Hubble constant.

The linear growth function \( D_0=D(a=1) \) is approximated using
the analytic form provided by \citet{Inman_2019}.
\[
D_{\mathrm{PBH}}(a) \approx \left(1 + \frac{3 \gamma a}{2 a_{-}
a_{\mathrm{eq}}} \right)^{a_{-}}, \quad \gamma =
\frac{\Omega_{\mathrm{m}} -
\Omega_{\mathrm{b}}}{\Omega_{\mathrm{m}}}
\]

\[
a_{-} = \frac{1}{4} \left( \sqrt{1 + 24 \gamma} - 1 \right),
\]
where \( a_{\mathrm{eq}} \approx \frac{1}{3400} \) denotes the
scale factor at the epoch of matter-radiation equality. At the
epoch of matter-radiation equality, Poisson fluctuations become
significant on scales \(k > k_{\rm eq} \approx
0.01\;\mathrm{Mpc}^{-1}\,.\)\citep{Placnck_2020}, And a
small-scale cutoff is applied with a step function \(
\Theta(k_{\mathrm{cut}} - k) \) for \( k > k_{cut} = \left( 2
\pi^2 n_{\text{PBH}}/{f_{\text{PBH}}} \right)^{1/3} \), following
\citet{Liu_2022}. The power spectrum in Eq.~(\ref{eq:P_tot}) is
normalized by matching to  \( \sigma(R = 8\, h^{-1}\mathrm{Mpc}) =
\sigma_8 \)\citep{Placnck_2020}.


In the case with extended mass spectrum, the mass fraction
function is correspondingly expressed as:
\[
f(m) = \frac{m}{ \Omega_{dm} \rho_{c}} \frac{dn}{dm}
\]
The relation between the mass of PBHs and the wavenumber of the
power spectrum can be naturally written as:
\[
m(k) \equiv 2\pi^{2} \Omega_{dm} \frac{\rho_{c}}{k^{3}}
\]
The influence of the mass function (\ref{eq:dn_dm_PBH}) of SMPBHs
originating from inflationary vacuum bubbles on the matter power
spectrum can be derived from Eq.\eqref{eq:P_PBH} as follows:
\begin{equation}
\label{eq:P_PBH_mf}
    \begin{aligned}
    P_{\mathrm{PBH}}(k)
    &= \int_{0}^{m(k)} \frac{f(m) m}{\Omega_{dm} \rho_{c}} D^{2} dm \\
    &= \frac{D^{2}}{\Omega_{dm}^{2} \rho_{c}^{2}} n_{0} \int_{0}^{m(k)} m^{2} \psi_{3}(m) dm \\
    &= \frac{D^{2}}{\Omega_{dm} \rho_{c}} \frac{f_{\mathrm{PBH}}}{M_{c}} e^{\sigma^{2}} \cdot \frac{M_{c}^{2} e^{-\sigma^{2}}}{2}\left[1+\operatorname{erf}\left(\frac{\ln \left(m / M_{c}\right)-\frac{\sigma^{2}}{2}}{\sqrt{2} \sigma}\right)\right] \\
    &= \frac{D^{2} f_{\mathrm{PBH}}}{2 \Omega_{dm} \rho_{c}} M_{c}\left[1+\operatorname{erf}\left(\frac{\ln \left(m / M_{c}\right)-\frac{\sigma^{2}}{2}}{\sqrt{2}
    \sigma}\right)\right].
    \end{aligned}
\end{equation}
Mathematically, it can be rigorously verified that
(\ref{eq:P_PBH_mf}) reduces to the monochromatic mass spectrum in
the limit $\sigma \to 0$.

\begin{figure}[htbp]
    \centering
    \begin{minipage}[b]{0.49\textwidth}
        \centering
        \includegraphics[width=\textwidth]{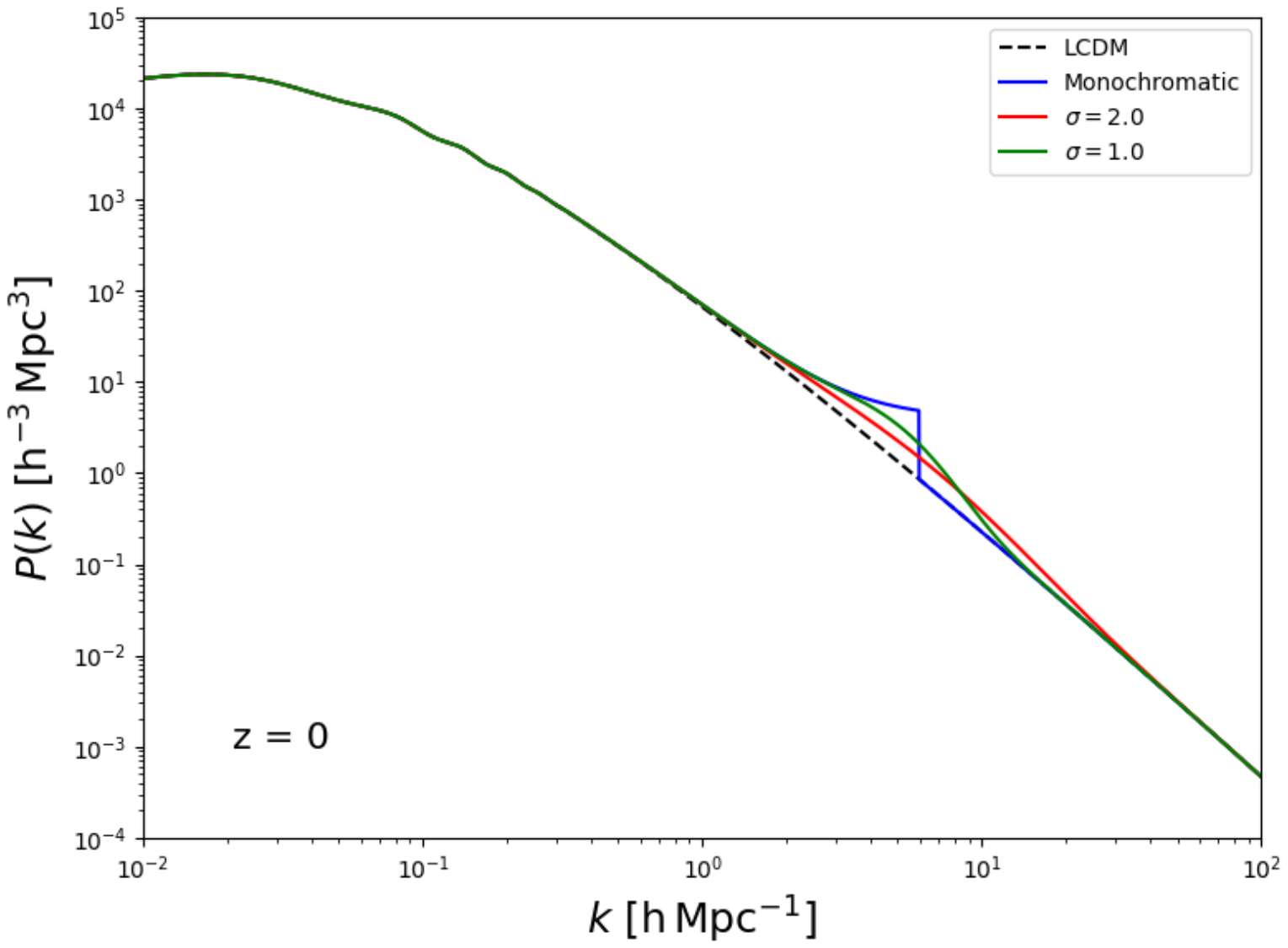}
    \end{minipage}
    \hfill
    \begin{minipage}[b]{0.49\textwidth}
        \centering
        \includegraphics[width=\textwidth]{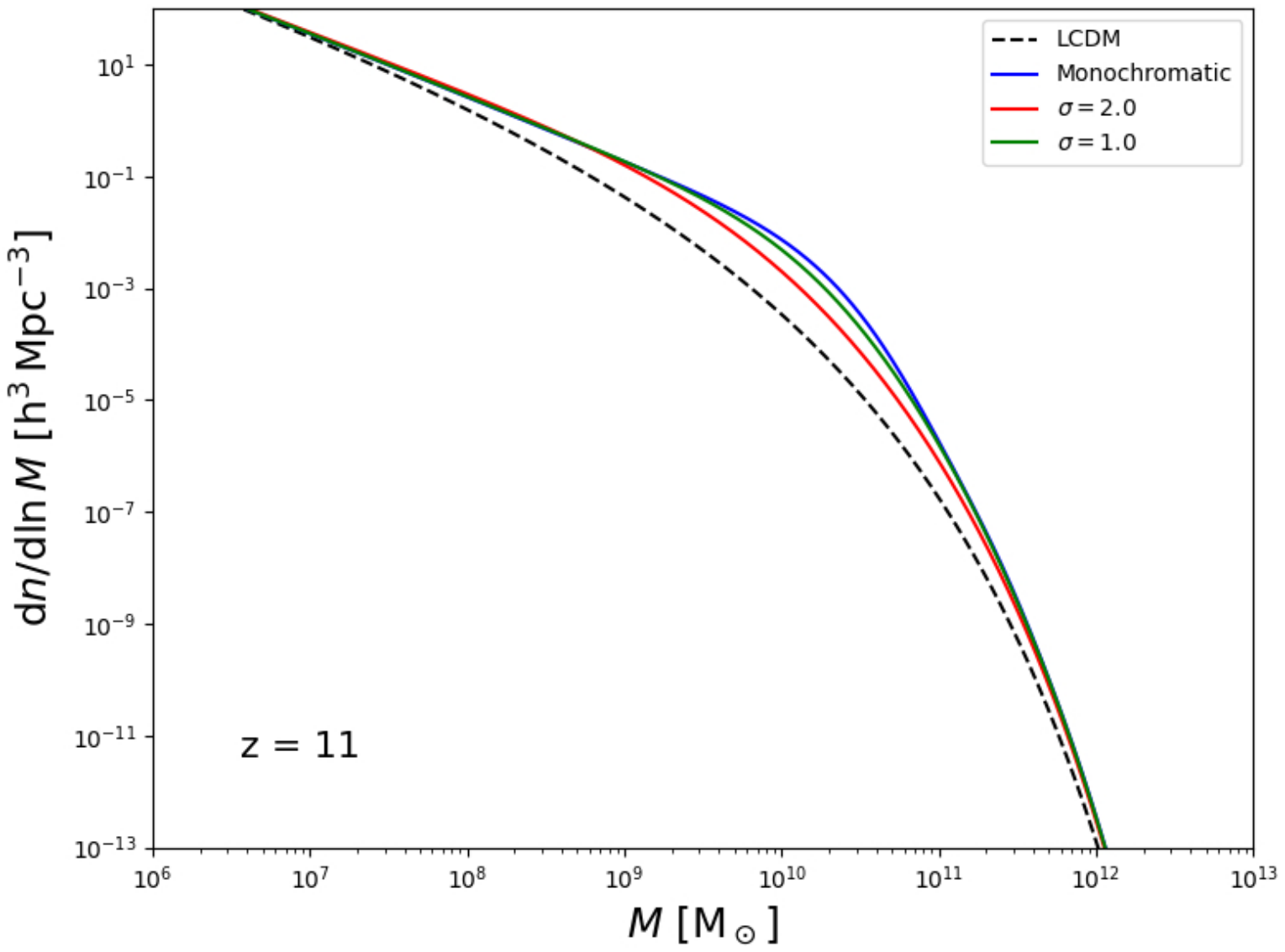}
    \end{minipage}
\caption{Left panel: Matter power spectra at \( z = 0 \) computed
for the standard \(\Lambda\)CDM model and for variants
incorporating
    SMPBHs with distinct mass distribution functions corresponding
    to different values of the parameter \(\sigma\)
    shown here for a specific PBH mass parameter
    \( M_{\mathrm{PBH}} = 10^{10}  \mathrm{M_{\odot}} \)
    and fractional abundance \( f_{\mathrm{PBH}} = 10^{-5} \).
    Right panel: corresponding HMF at \(z=11\).
    The black dashed line shows the $\Lambda$CDM result.
    The blue line corresponds to the inclusion of SMPBHs
    with a monochromatic mass function,
    while the orange and green lines represent
    those with our extended mass functions (\ref{eq:dn_dm_PBH}) with width parameters
    $\sigma=2.0$ and $\sigma=1.0$, respectively.
    }
    \label{fig:2}
\end{figure}


\bibliography{refs-2}{}



\end{document}